\documentclass[sigconf]{acmart}
\usepackage{fancyhdr}
\usepackage[T1]{fontenc}
\usepackage[utf8]{inputenc}
\usepackage{microtype}
\usepackage{ragged2e}

\usepackage{hyperref}
\hypersetup{
    unicode=true,
    pdfencoding=auto,
    pdftitle={Listen to Rhythm, Choose Movements: Autoregressive Multimodal Dance Generation via Diffusion and Mamba with Decoupled Dance Dataset},
    pdfauthor={}
}

\renewcommand\footnotetextcopyrightpermission[1]{}
\acmConference{}{}{}
\acmYear{}
\acmBooktitle{}
\acmISBN{}
\acmDOI{}       

\begin{document}
\raggedbottom

\title{Listen to Rhythm, Choose Movements: \\
 Autoregressive Multimodal Dance Generation via \\
 Diffusion and Mamba with Decoupled Dance Dataset}

\author{Oran Duan}
\authornote{Work was done when interned at Zhipu AI.}
\affiliation{
  \institution{
    Communication University of China
  }
  \city{Beijing}
  \country{China}
}
\email{oranduan@cuc.edu.cn}

\author{Yinghua Shen}
\affiliation{
  \institution{
    Communication University of China
  }
  \city{Beijing}
  \country{China}
}
\email{shenwan@cuc.edu.cn}

\author{Yingzhu Lv}
\affiliation{
  \institution{
    Communication University of China
  }
  \city{Beijing}
  \country{China}
}
\email{bamboo_lv@mails.cuc.edu.cn}

\author{Luyang Jie}
\authornote{Corresponding author.}
\affiliation{
  \institution{
    Zhipu AI
  }
  \city{Beijing}
  \country{China}
}
\email{luyang.jie@aminer.cn}

\author{Yaxin Liu}
\affiliation{
  \institution{
    Zhipu AI
  }
  \city{Beijing}
  \country{China}
}
\email{yaxin.liu@aminer.cn}

\author{Qiong Wu}
\affiliation{
  \institution{
    Zhipu AI
  }
  \city{Beijing}
  \country{China}
}
\email{qiong.wu@aminer.cn}

\begin{abstract}
Advances in generative models and sequence learning have greatly promoted research in dance motion generation,  
yet current methods still suffer from coarse semantic control and poor coherence in long sequences.  
In this work, we present \textbf{Listen to Rhythm, Choose Movements} (LRCM),  
a multimodal-guided diffusion framework supporting both diverse input modalities and autoregressive dance motion generation.  
We explore a feature-decoupling paradigm for dance datasets and generalize it to the Motorica Dance dataset,  
separating motion capture data, audio rhythm, and professionally annotated global and local text descriptions.  
Our diffusion architecture integrates an audio–latent Conformer and a text–latent Cross-Conformer,  
and incorporates a Motion Temporal Mamba Module (MTMM) to enable smooth, long-duration autoregressive synthesis.  
Experimental results indicate that LRCM delivers strong performance in both functional capability and quantitative metrics,  
demonstrating notable potential in multimodal input scenarios and extended-sequence generation.  
The project page is available at \url{https://oranduanstudy.github.io/LRCM/}.
\end{abstract}

\begin{teaserfigure}
  \centering
  \includegraphics[width=\textwidth]{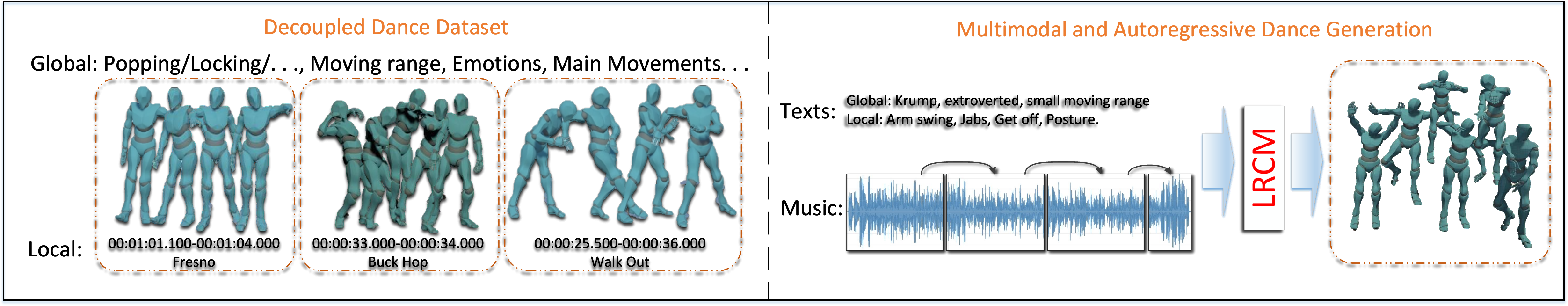} % Replace with the path to your image
  \vspace{-2.0em} % Adjust vertical space as needed
  \caption{A conceptual overview of the proposed LRCM framework, showing the decoupling of the text modality from the Motorica Dance dataset, integrated into a multimodal autoregressive dance generation model based on the diffusion structure.}
  \label{fig:overview}
  \Description{Illustration showing modality decoupling (motion, audio, text) and multimodal-guided diffusion architecture for autoregressive dance generation.}
\end{teaserfigure}

\maketitle
\section{Introduction}

The task of \textit{dance motion generation} aims to automatically produce sequences of dance movements that are highly synchronized with a given music track. Such generation requires precise alignment across rhythm, melody, 
and stylistic dimensions, while ensuring that the dynamical visual expression and emotional intent are deeply integrated with the music content~\cite{zhu2023human}. Generative models have achieved remarkable advances 
in various domains---including image synthesis, text generation, speech synthesis, and video generation---and have also demonstrated strong potential in motion synthesis and other sequence generation tasks. By analyzing key features 
from the musical signal, such as beats, intensity, and tonality, a generative motion model should effectively produce spatial-temporal movement trajectories and amplitudes that accurately align the dancer's motion with musical variations.

With the evolution of generative architectures, the field of motion generation has gradually shifted towards higher-capacity models capable of richer expressiveness. Recent methods such as Variational Autoencoders (VAE)~\cite{biquard2025variational} and 
Vector Quantized VAE (VQ-VAE)~\cite{qi2025human} have become hotspots in the domain. While VAEs model the probabilistic distribution of the latent space to explore motion diversity, VQ-VAEs further employ a discrete codebook to enhance structural 
representation. Meanwhile, Diffusion Models (DM)~\cite{croitoru2023diffusion} and Latent Diffusion Models (LDM)~\cite{rombach2022high} have emerged as powerful generative frameworks for high-dimensional motion data, introducing iterative denoising and 
fine-grained sampling to approach real data distributions closely. These models significantly improve motion fidelity and structural detail, and effectively capture complex spatio-temporal dependencies of human movement.

However, the continued development of generative models poses new challenges, especially with the diversification and expansion of datasets. In dance generation tasks, semantic information has evolved from unimodal input to complex 
multimodal systems integrating text, audio, music, and contextual scene data. Early studies often relied on simple mappings between a single modality and motion (e.g., text-to-motion or audio-to-motion), whereas modern applications demand 
integration of rich multimodal semantics. This trend, together with the rise of large models, underscores the cross-disciplinary need for unified, generalizable motion generation models. Moreover, the introduction of multiple modalities imposes 
substantial challenges in efficiency, scalability, and precision—especially for deployment under regular hardware constraints. Furthermore, long-sequence generation remains difficult to reconcile between fidelity and efficiency. Breakthroughs in 
long-sequence modeling in computer vision and NLP motivate opportunities to adapt these advances for dance generation, potentially enabling higher-quality and more practical applications.

Motivated by these gaps, we focus on key research problems in dance generation and introduce a novel framework leveraging the Motorica Dance motion capture dataset. Our contributions include:
\begin{itemize}
    \item \textbf{Decoupled Multimodal Dance Dataset Paradigm:}  
    We propose a fine-grained semantic decoupling paradigm for multimodal dance datasets,  
    which formalizes the separation of dance motion, audio rhythm, and professionally annotated textual descriptions  
    into hierarchical global style and local movement levels.  
    This paradigm is instantiated and validated on the Motorica Dance dataset,  
    enabling systematic representation of multimodal dance data and supporting subsequent experiments on text-guided motion generation.

    \item \textbf{Heterogeneous Multimodal-Guided Diffusion Architecture:}  
    We design a diffusion-based framework that asymmetrically fuses audio and text modalities.  
    Audio–latent conformers are employed to capture persistent rhythmic cues,  
    while text–latent cross-conformers incorporate fine-grained semantics from global and local textual inputs.  
    A jerk-based loss function is introduced to jointly maintain rhythmic smoothness and semantic consistency.
    
    \item \textbf{State Space Model-based Autoregressive Temporal Module:}  
    We explore an autoregressive extension of the diffusion framework for long-sequence generation  
    by integrating the Motion Temporal Mamba Module (MTMM), built upon state space modeling principles.  
    This component is designed to efficiently capture long-range motion dependencies,  
    supporting experiments on generation diversity and temporal continuity in extended dance sequences.
\end{itemize}
We term our complete pipeline the \textbf{Listen to Rhythm, Choose Movements} (\textbf{LRCM}) framework,  
a multimodal dance generation paradigm that combines audio rhythm and textual semantics in a dual-conditioning scheme.  
Designed to provide fine-grained and context-aware control over motion synthesis across both short and long time scales,  
LRCM integrates feature-decoupled dataset representations with a heterogeneous diffusion architecture and autoregressive temporal modeling.  
An overview of its components and workflow is illustrated in Figure~\ref{fig:overview}.

%% --------Related Work--------
\section{Related Work}

\subsection{Motion Representation and Datasets}

Human motion data is commonly represented as temporal sequences of body poses.  
Existing research generally adopts either \textit{keypoint-based} representations 
or \textit{rotation-based} formats~\cite{ma20233d,zanfir2021thundr}.  
More recent studies explore \textit{statistical mesh models}~\cite{zheng2023deep} for increased accuracy,  
with the Skinned Multi-Person Linear Model (SMPL)~\cite{loper2023smpl} being a representative approach  
for modeling body shapes and surface features.  

Datasets for motion generation and dance synthesis have diversified over time.  
For general human motion generation tasks, core datasets such as HumanML3D~\cite{guo2022generating}  
and KIT~\cite{plappert2016kit} use structured, detailed textual labels to annotate human motion clips.  
In contrast, dance motion datasets emphasize musical and artistic characteristics.  
Representative examples include AIST++~\cite{li2021ai}, FineDance~\cite{li2023finedance},  
and MotoricaDance~\cite{valle2021transflower}, whose primary annotations consist of  
music segments with corresponding dance motion sequences.  
Compared to general motion datasets, textual descriptions in dance datasets are typically coarse,  
often reduced to style tags or simple notes about music attributes, lacking systematic,  
fine-grained linguistic representations.  
This limitation poses challenges for text-driven dance modeling and multimodal fusion requiring precise semantic control.

\subsection{3D Motion Generation}

Single-modality motion-driven algorithms have matured considerably,  
particularly within architectures built upon Variational Autoencoders (VAEs) and Diffusion Models.  
In the VAE-based generation domain, for example,  
T2M-GPT~\cite{zhang2023generating} combines a pretrained GPT for text-driven motion synthesis.  
Similarly, Momask~\cite{guo2024momask} employs residual VQ-VAE to enhance text-to-motion generation,  
while Bailando~\cite{li2022bailando} extends the VQ-VAE approach for music-to-dance mapping.  
In parallel, diffusion-based motion generation has experienced rapid growth  
and has been widely adopted for both general motion and dance tasks.  
Notably, MDM~\cite{tevet2023human} is among the first to apply diffusion models to motion data.  
Building upon this, EDGE~\cite{tseng2023edge} designs an editable diffusion framework for dance synthesis.  
Lodge~\cite{li2024lodge} advances the field by applying multi-level granularity control  
for both coarse- and fine-style dance actions,  
and LGTM~\cite{sun2024lgtm} integrates CLIP with diffusion to support multi-granularity text-motion fusion.  

When moving toward multimodal and hierarchical conditioning for dance generation,  
several works have explored combining diverse modalities.  
TM2D~\cite{gong2023tm2d}, for example, achieves preliminary fusion  
between dance-text and music-motion information.  
LM2D~\cite{yin2024lm2d} further incorporates both lyrics and music for joint conditioning,  
while Beat-It~\cite{huang2024beat} focuses specifically on the influence of drum beats and rhythm in motion generation.  
In addition, LMM~\cite{zhang2024large} explores multi-dataset integration  
to support multi-task dance generation,  
and GCDance~\cite{liu2025gcdance} investigates using category-based music input for motion editing.  
Despite these advancements, achieving deep and efficient alignment across modalities  
remains a core challenge for the field.

\subsection{Long-Sequence Modeling Tasks}

Long-sequence modeling has gained significant attention in recent deep learning research.  
Early methods include Recurrent Neural Networks (RNNs)~\cite{lipton2015criticalreviewrecurrentneural} and Long Short-Term Memory (LSTMs)~\cite{lindemann2021survey},  
which can capture temporal dependencies across frames  
but suffer from gradient vanishing and difficulty in handling extended sequences.  
For example, JL2P~\cite{ahuja2019language2pose} introduced a gated recurrent unit (GRU) text encoder  
to align language and motion features;  
Ma~et~al.~\cite{ma2022DanceAction} trained a hybrid-density LSTM on modern dance motion-capture data;  
while Nelson Yalta~et~al.~\cite{Nelson2019WeaklySupervised} explored weakly-supervised real-time dance generation  
using hybrid-LSTM architectures.
Transformer-based methods~\cite{vaswani2017attention} introduced self-attention to capture global dependencies,  
leading to breakthroughs in NLP and CV, and gaining popularity in dance generation.  
However, the quadratic complexity of self-attention limits scalability, making long-sequence training and inference costly.

Recently, Mamba~\cite{gu2023mamba} has emerged as an efficient state space model (SSM) architecture  
with linear complexity and hardware-friendly design,  
ideal for long-sequence tasks.  
MotionMamba~\cite{zhang2024motion} pioneered applying Mamba to motion generation,  
enabling long-span dance synthesis;  
MatchDance~\cite{yang2025matchdance} achieved SOTA results on FineDance   
by proposing a Mamba-Transformer hybrid;  
MambaDance~\cite{park2025not} replaced self-attention entirely with Mamba  
for high-precision music-dance synchronization;  
MEGADance~\cite{yang2025MEGADance} introduced Mixture-of-Experts (MoE) mechanisms  
to improve style continuity and beat alignment.  
Additionally, architectures like TTT~\cite{sun2025ttt} further expand design choices for sequence modeling,  
with hybrid innovations steadily advancing controllable long-sequence motion synthesis.

%% --------Methodology--------
\section{Method}

\subsection{Preliminary}
\paragraph{Conditional Diffusion Models.}  
The primary objective of our neural network is to train a denoising diffusion generative model.  
Given a music waveform and a textual description, we encode them into conditional embedding vectors  
$c_a$ and $c_t$.  
These embeddings are injected into the diffusion network to generate a latent motion sequence  
$x_{1:T} \in \mathbb{R}^{B \times L \times J}$,  
which is subsequently decoded by a motion decoder into a dance motion sequence usable in deployment scenarios.

Let $x$ follow an unknown density distribution $q(x)$.  
To construct the diffusion model of $x$, we first define a diffusion process  
as a Markov chain $q(x_n \mid x_{n-1})$ for $n \in \{1, \ldots, N\}$.  
During training, Gaussian noise is progressively added to the observation $x_0$ ($n=0$)  
until it is fully destroyed,  
so that $q(x_n \mid x_0)$ follows a standard normal distribution.  
The idea is to train a network to invert the $q$ process,  
reversing the diffusion steps to recover an observation from noise.  
This produces the reverse (denoising) process $p$.  
Assuming the noise at each step is zero-mean Gaussian, we have:
\begin{equation}
q(x_n \mid x_{n-1}) = \mathcal{N}(x_n; \alpha_n x_{n-1}, \beta_n I)
\label{eq:forward_diff}
\end{equation}
Here, $\mathcal{N}(\cdot; \mu, \Sigma)$ denotes a multivariate Gaussian distribution at $x_n$  
with mean $\mu = \alpha_n x_{n-1}$ and covariance matrix $\Sigma = \beta_n I$.  
By setting $\alpha_n = \sqrt{1 - \beta_n}$,  
the set $\{\beta_n\}_{n=1}^N$ fully specifies the diffusion process~\cite{ho2020denoising}.  
If the noise magnitude at step $n$ is small relative to $x_n$,  
the reverse distribution can also be approximated by a Gaussian~\cite{beyerle2024inferring}.  
This Gaussian approximation can be written as:
\begin{equation}
p(x_n) = \mathcal{N}(x; 0, I)
\label{eq:prior}
\end{equation}
\begin{equation}
p(x_{n-1} \mid x_n) = \mathcal{N}(x_{n-1}; \mu(x_n, n), \Sigma(x_n, n))
\label{eq:reverse_diff}
\end{equation}

Training proceeds via score matching, similar to energy-based models.  
The loss function can be expressed as:
\begin{equation}
\mathcal{L}_{diff} = \mathbb{E}_{x_0, n, \epsilon}
\left[ \kappa_n \, \left\| \epsilon - \hat{\epsilon}(\widetilde{\alpha_n} x_0 + \widetilde{\beta_n} \epsilon, n) \right\|_2^2 \right]
\label{eq:diff_loss}
\end{equation}
Here, $x_0$ is uniformly sampled from the training set $\mathcal{D}$,  
$n \in \{1, \ldots, N\}$, and $\epsilon \sim \mathcal{N}(0, I)$.  
$\hat{\epsilon}(x, n)$ denotes the network’s prediction of the noise added to $x_0$.  
$\widetilde{\alpha_n}$ and $\widetilde{\beta_n}$ are constants determined by $\{\beta_n\}$,  
and $\kappa_n$ is a weight (often set to 1 for better subjective performance).  
A secondary term $- \log p(x_0 \mid x_1)$, based on negative log-likelihood, is also included.

For multi-conditional inputs (e.g., audio and text),  
the task becomes training a conditional diffusion model $\hat{\epsilon}(x; c_a, c_t; n)$.  
The relative influence of each condition is controlled via guided diffusion~\cite{dhariwal2021diffusion},  
which modifies the reverse process as:
\begin{equation}
p_\gamma(x_{n-1} \mid x_n; c_a, c_t) 
\propto p(x_{n-1} \mid x_n) \, p(c_a \mid x_n)^\gamma \, p(c_t \mid x_n)^\delta
\label{eq:guided}
\end{equation}
where $\gamma, \delta > 0$ are scaling coefficients for conditional strength.

\vspace{0.5em}
\paragraph{Mamba and State Space Models (SSMs).}  
For long-sequence modeling,  
state space models (SSMs)---especially recent developments like S4 and Mamba---  
have proven effective~\cite{gu2023mamba}.  
These models process a one-dimensional input sequence $x(t) \in \mathbb{R}$ into an output $y(t) \in \mathbb{R}$  
via a hidden state $h(t) \in \mathbb{R}^N$.  
They define trainable evolution parameters $A \in \mathbb{R}^{N \times N}$,  
input projection $B \in \mathbb{R}^{N \times 1}$,  
and output projection $C \in \mathbb{R}^{1 \times N}$.  
The continuous-time dynamics are:
\begin{equation}
h'(t) = A\, h(t) + B\, x(t)
\label{eq:ssm_cont1}
\end{equation}
\begin{equation}
y(t) = C\, h(t)
\label{eq:ssm_cont2}
\end{equation}

To enable practical computation,  
S4 and Mamba discretize these state space equations.  
Given discretization functions $f_A(\Delta, A)$ and $f_B(\Delta, A, B)$,  
a common choice is zero-order hold (ZOH), yielding:
\begin{equation}
\overline{A} = \exp(\Delta A), \quad
\overline{B} = (\Delta A)^{-1} \left( \exp(\Delta A) - I \right) \cdot \Delta B
\label{eq:discretization}
\end{equation}
After discretization $(\Delta, A, B, C) \mapsto (\overline{A}, \overline{B}, C)$,  
the model becomes:
\begin{equation}
h_t = \overline{A} \, h_{t-1} + \overline{B} \, x_t
\label{eq:ssm_disc1}
\end{equation}
\begin{equation}
y_t = C\, h_t
\label{eq:ssm_disc2}
\end{equation}
For efficient output computation,  
a structured 1D convolution kernel $\overline{K}$ is precomputed:
\begin{equation}
\overline{K} = \left( C \overline{B}, \; C \overline{A} \overline{B}, \; \ldots, \; C \overline{A}^{L-1} \overline{B} \right)
\label{eq:kernel}
\end{equation}
\begin{equation}
y = x * \overline{K}
\label{eq:conv_output}
\end{equation}

The Mamba model incorporates hardware-aware parallel scanning algorithms,  
integrating the selective SSM into an end-to-end neural network,  
and thus enabling efficient, scalable long-sequence modeling.

\subsection{Decoupled Dance Dataset Paradigm}
The spatiotemporal interaction between dance and music forms a fundamental basis for research in dance generation.  
Text, as another critical modality, is often overlooked in existing tasks.  
From the perspective of cross-modal semantic association,  
dance-related textual descriptions not only convey artistic style and motion semantics,  
but also provide high-level semantic cues that purely motion and audio signals cannot capture~\cite{li2023labanformer,ikeuchi2018describing,brown2024performing}.  
These textual annotations, when integrated with other modalities, enable more precise semantic conditioning and richer generative control.

Motivated by the need for systematic handling of multimodal data,  
we introduce a \textbf{feature-decoupling paradigm} for dance datasets.  
This paradigm formalizes the separation of motion capture sequences $A$,  
audio signals $S$, and textual descriptions obtained via professional dance artistry annotation,  
ensuring that each modality is independently modeled while preserving semantic consistency across modalities.  
The decoupling process treats these components as carriers of complementary motion semantics and organizes them  
into a structured, interpretable format suitable for downstream generative modeling.

\begin{figure}[ht]
    \centering
    \includegraphics[width=\linewidth]{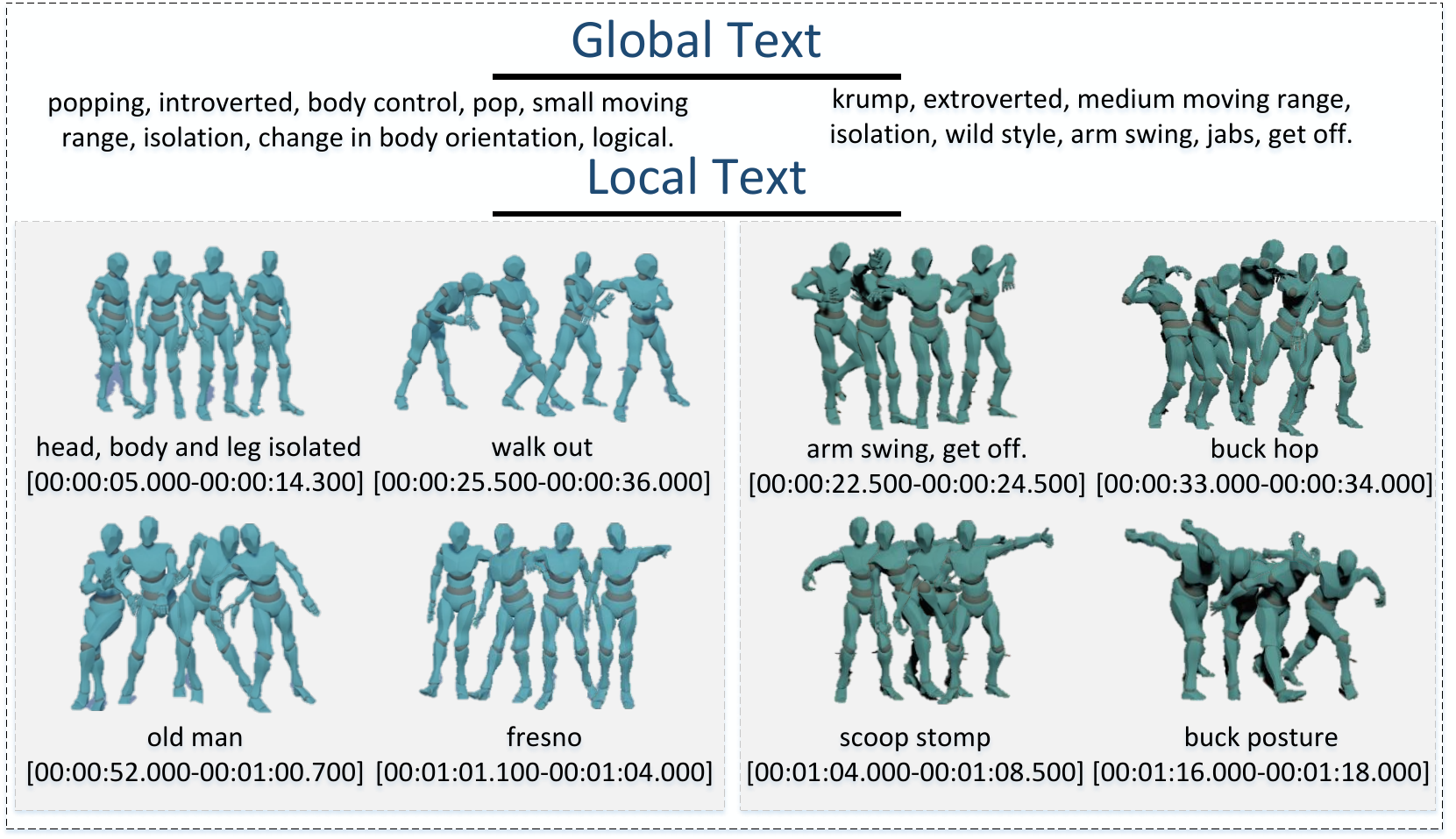}
    \vspace{-2.0em}
    \caption{Example of decoupled textual modality annotations within the proposed paradigm.}
    \label{fig:text_annotation}
    \Description{Visualization of global and local text tags associated with annotated dance clips.}
\end{figure}

From a dataset engineering standpoint,  
the paradigm employs motion feature modeling $f_{motion}$ and audio feature modeling $f_{audio}$  
to generate representation vectors from raw motion and audio inputs:
\begin{equation}
F^{motion}_i = f_{motion}(A_i), \quad F^{audio}_i = f_{audio}(S_i)
\label{eq:motion_audio}
\tag{12}
\end{equation}
Furthermore, an engineered artistic decoupling method extracts professional textual features from the motion data:
\begin{equation}
F^{text}_i = Decouple(A_i)
\label{eq:text}
\tag{13}
\end{equation}
Finally, the multimodal decoupling paradigm represents each sample in a fully structured three-modality form:
\begin{equation}
D' = \{ (F^{motion}_i, F^{audio}_i, F^{text}_i) \}_{i=1}^N
\label{eq:three_modality}
\tag{14}
\end{equation}

We generalize this paradigm to the \textbf{Motorica Dance dataset},  
applying the same decoupling process to produce motion–audio–text triples.  
To ensure annotation quality, seven professional street dancers contributed to the textual labeling.  
Annotations were organized into two levels:  
\textit{global labels}, providing coarse-grained descriptions of dance category, spatial displacement range, artistic style, and core movements;  
and \textit{local labels}, capturing fine-grained characteristics and variations of specific movements over time.  
This hierarchical design supports comprehensive separation of rhythm, motion, and semantics,  
providing fine-grained conditional control for subsequent multimodal and autoregressive generative tasks.

An example of the decoupled textual annotation, adhering to the proposed paradigm, is shown in Figure~\ref{fig:text_annotation}.

\subsection{Main Model Architecture}
\begin{figure*}[t]
    \centering
    \includegraphics[width=\textwidth]{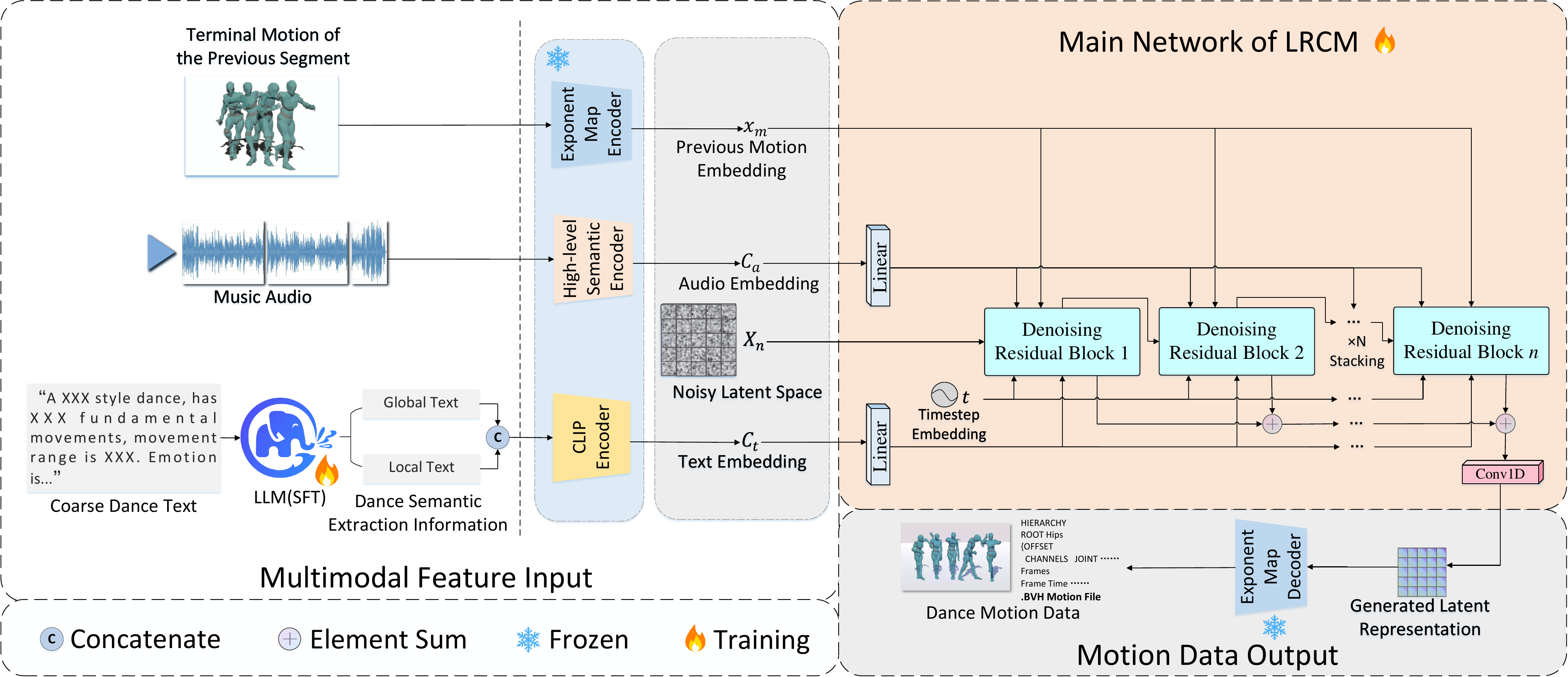} % 宽度跨两列
    \vspace{-2.0em}
    \caption{Overview of the proposed architecture.  
    The main generation model is constructed using a DiT backbone.  
    Text inputs are first processed through a fine-tuned LLM performing semantic tokenization,  
    standardizing informal or non-standard prompts;  
    the resulting tokens are then passed to a CLIP-based text encoder.  
    Audio features are extracted via a high-level semantic encoder.  
    The model stacks $N$ denoising residual blocks with residual and skip connections  
    to integrate the encoded audio and text features for dance motion generation.}
    \Description{Large diagram showing backbone network with audio encoder, text encoder, and stacked denoising residual blocks.}
    \label{fig:fig3_architecture}
\end{figure*}
\paragraph{Multimodal-guided dance generation.}
Given a set of multimodal audio embeddings  
$C_a \in \mathbb{R}^{T_{\text{seq}} \times D_a}$  
and text embeddings  
$C_t \in \mathbb{R}^{T_{\text{seq}} \times D_t}$,  
our goal is to train a dance-generation diffusion model  
such that, for noise steps $t \in [0, \text{timesteps}]$,  
the network minimizes the distance between the predicted noise  
and true Gaussian noise on the conditioned input features.  

After multi-step denoising,  
the model outputs motion representation  
$P \in \mathbb{R}^{T_{\text{seq}} \times D_p}$.  
The motion representation is encoded using exponent-map-based pose encoding,  
while audio representations are obtained via the \texttt{librosa} high-level semantic audio processing library.  

Here, $T_{\text{seq}}$ denotes the sequence length,  
$D_a = 3$ and $D_p = 61$ are the respective feature dimensions.
Text features are embedded into the denoising network  
using a pretrained CLIP encoder (ViT-B/32),  
with dimension $D_t = 512$.  
Throughout the network,  
encoder–decoder components use frozen pretrained weights.

As shown in Figure~\ref{fig:fig3_architecture},  
our primary network architecture is inspired by the single-modality diffusion model of LDA~\cite{alexanderson2023listen},  
which uses DiT with skip connections and stacked residual layers.  
It operates in a parallel, non-causal manner,  
allowing non-causal convolution when generating dance from a music segment.

For text inputs,  
we finetune a \texttt{GLM-4-FLASH} pretrained large language model~\cite{glm2024chatglm}  
to refine user-provided non-professional dance descriptions  
into standardized textual prompts usable by the network.  
Texts are further separated into \textit{global} and \textit{local} descriptions. 
More details on text processing are provided in Appendix~\ref{sec:llm_finetune}

\begin{figure}[t]
    \centering
    \includegraphics[width=\linewidth]{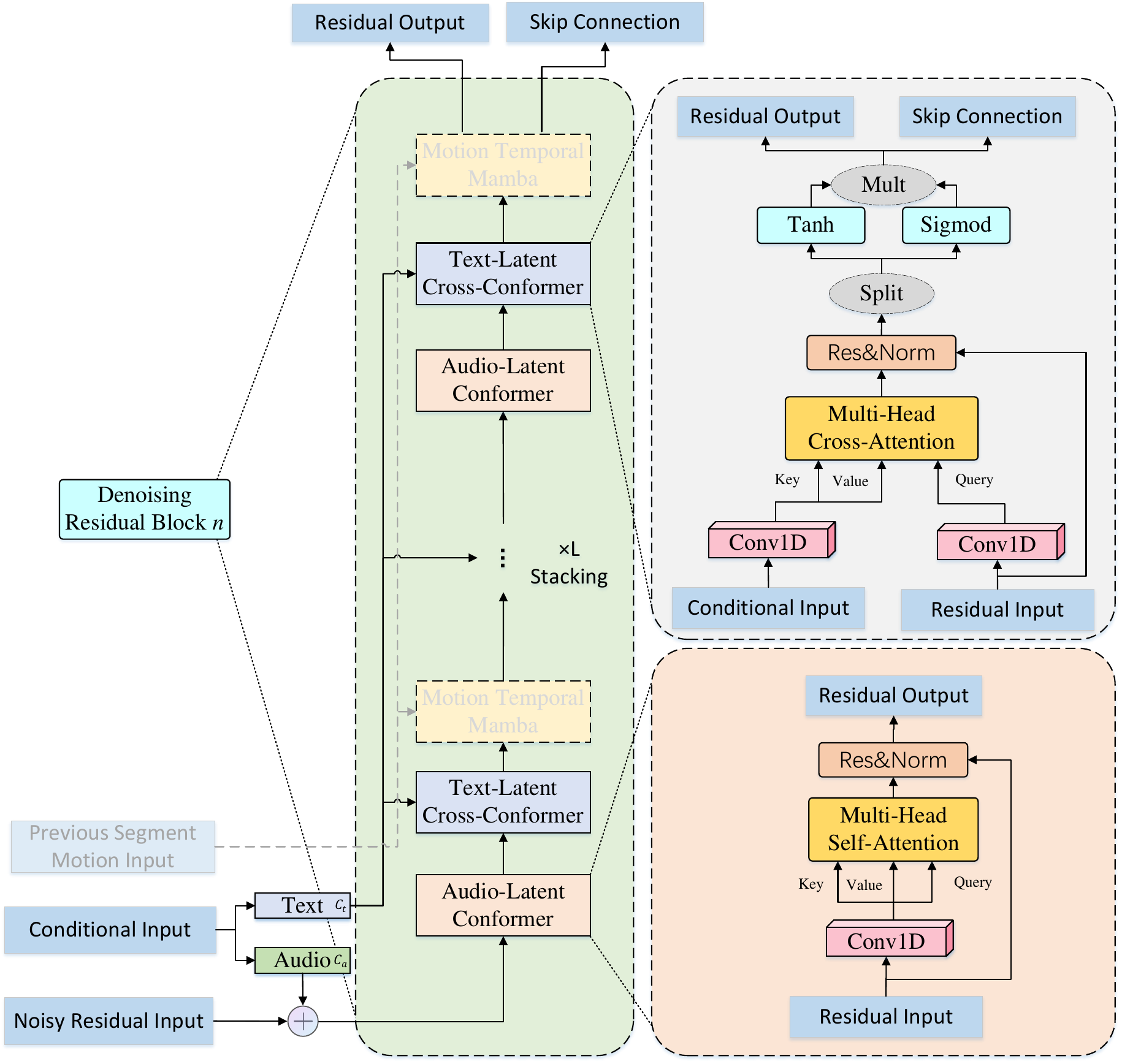}
    \vspace{-2.0em} 
    \caption{
        Internal structure of the denoising residual block.  
        Each block stacks $L$ computation layers following the Conformer architecture,  
        applying audio-specific and text-specific processing respectively.  
        Two conditional embedding modes are employed:  
        \textit{Self-Attention} for intra-modal feature refinement,  
        and \textit{Cross-Attention} for cross-modal integration.  
        A Motion Temporal Mamba module is appended after the attention layers  
        to capture long-range temporal dependencies.
    }
    \Description{
        Diagram showing the residual denoising block,  
        with audio–latent Conformer, text–latent cross-Conformer,  
        multi-head self-attention and cross-attention units,  
        and a Motion Temporal Mamba module at the end  
        for modeling long temporal dependencies in dance sequences.
    }
    \label{fig:fig4_residual_block}
\end{figure}

\paragraph{Core Diffusion Model and Loss Functions.}
The main diffusion model is composed of stacked denoising residual blocks.  
As shown in Figure~\ref{fig:fig4_residual_block},  
each residual block contains an audio–latent Conformer,  
a text–latent cross-Conformer,  
and a motion-temporal Mamba module (explained in Section~3.4).  

Audio features are injected in a manner similar to the original backbone:  
after feature projection,  
they are linearly added to the latent noise sequence $X_n$ representing motion poses.  
In the first audio–latent Conformer,  
the latents are enhanced via 1D convolution,  
processed by multi-head self-attention (MHSA),  
and then passed through residual and layer normalization (LN).  
This can be expressed as:
\begin{equation}
x_a = \mathrm{LN}\left( X_n \oplus \mathrm{MHSA}\left( \mathrm{Conv1D}\left( C_a \oplus X_n \right) \right) \right)
\label{eq:audio_conformer}
\end{equation}
Here, MHSA denotes multi-head self-attention,  
LN denotes layer normalization,  
and $x_a$ is the latent output from the audio–latent Conformer block.

Text, as the second cross-modal embedding,  
is introduced using a different scheme from audio,  
because we consider music to be the dominant modality driving dance generation,  
while text provides semantic content to structure dance movement.  

We design a text–latent cross-Conformer module:  
the input consists of global and local text embeddings,  
which are downsampled via bottleneck networks,  
activated, concatenated, and projected into a joint text feature $C_t$.  
After 1D convolution enhancement,  
$C_t$ is projected into Key and Value matrices,  
while $x_a$ from the audio module (also Conv1D-enhanced) serves as the Query.  
Multi-head cross-attention (MHCA) is then applied,  
followed by residual connection and layer normalization:
\begin{equation}
x_t = \mathrm{LN}\left( x_a \oplus \mathrm{MHCA}\left( \mathrm{Conv1D}(x_a), \mathrm{Conv1D}(C_t) \right) \right)
\label{eq:text_conformer}
\end{equation}
We employ a Gated Tanh Unit (GTU) as the activation post-processing step.  
The latents are expanded in dimension,  
split into gating $h_g$ and filtering $h_f$ components,  
and activated via $\tanh(\cdot)$ and $\sigma(\cdot)$:  
\begin{equation}
h_g, h_f = \mathrm{Split}\left( \mathrm{Expand}(x_t), 2, \mathrm{dim}=2 \right)
\label{eq:gtu_split}
\end{equation}
\begin{equation}
X_{n^\prime}, h_{\mathrm{skip}} = \mathrm{Split}\left( \tanh(h_f) \odot \sigma(h_g) \right)
\label{eq:gtu_merge}
\end{equation}

Here, $h_{\mathrm{skip}}$ denotes the latent skip connection to be linearly added in later stages,  
$\mathrm{Split}(\cdot)$ indicates tensor splitting,  
$\mathrm{Expand}(\cdot)$ indicates dimensional expansion,  
and $X_{n^\prime}$ is the latent input to the next residual block.  
In the non-autoregressive framework,  
$X_{n^\prime}$ can be directly passed to the next denoising block.

\paragraph{Motion Reconstruction Losses.}  
To encourage accurate reconstruction of temporal motion dynamics,  
we design three derivative-based loss components: velocity, acceleration, and jerk.  
All derivatives are normalized by their respective L2 norms  
(with a small constant $\varepsilon > 0$ added to ensure numerical stability)  
to focus on the variation pattern rather than magnitude.

The generic form of the motion derivative loss is:
\begin{equation}
\mathcal{L}_{\mathrm{motion}}^{q} =
\sum_{d=1}^{D_p}
\left(
\frac{q_{d}^{\mathrm{pred}}}{\max\left( \| \mathbf{q}^{\mathrm{pred}} \|_2, \varepsilon \right)}
-
\frac{q_{d}^{\mathrm{gt}}}{\max\left( \| \mathbf{q}^{\mathrm{gt}} \|_2, \varepsilon \right)}
\right)^2
\label{eq:motion_loss_core}
\end{equation}
Here $q^{\mathrm{pred}}$ and $q^{\mathrm{gt}}$ denote the predicted and ground-truth derivative vectors,  
$q$ indexes over first-order (\textbf{velocity}), second-order (\textbf{acceleration}), and third-order (\textbf{jerk}) derivatives,  
and $D_p$ is the pose feature dimension.
The total training objective is defined as:
\begin{equation}
\mathcal{L}_{total} =
\mathcal{L}_{diff} +
\lambda_v \, \mathcal{L}_{\mathrm{vel}} +
\lambda_a \, \mathcal{L}_{\mathrm{acc}} +
\lambda_j \, \mathcal{L}_{\mathrm{jerk}}
\label{eq:total_loss}
\end{equation}

\subsection{Motion Temporal Mamba Module}
\begin{figure*}[t]
    \centering
    \includegraphics[width=\textwidth]{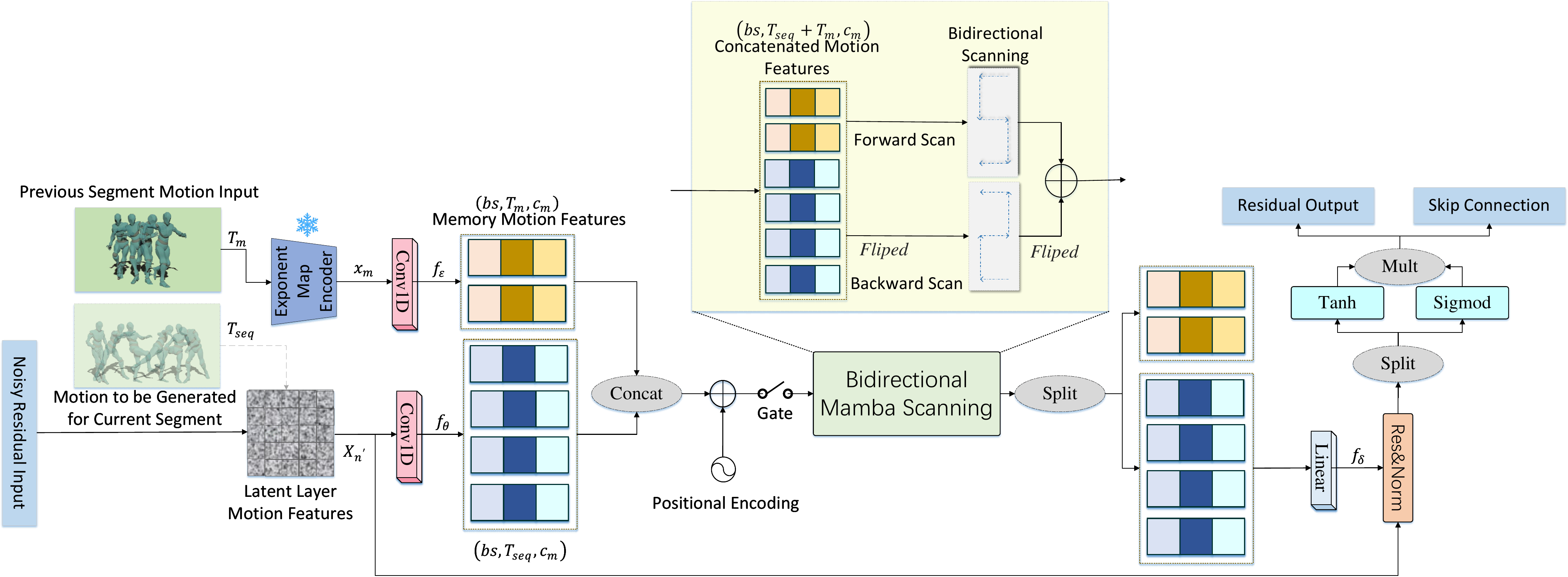}
    \vspace{-2.0em}
    \caption{
        Motion Temporal Mamba Module (MTMM) architecture and process.  
        Latent features from past motion memory and current motion are concatenated along the temporal axis,  
        enriched with positional encoding,  
        and processed by a bidirectional Mamba scan.  
        The output is split to obtain the current segment's latent motion features for subsequent generation.
    }
    \Description{
        Diagram of MTMM: inputs from past motion memory and the current segment are concatenated,  
        positional encoding is applied,  
        a forward and backward Mamba scan is performed,  
        and resulting features are split to form outputs for the next stage in autoregressive generation.
    }
    \label{fig:fig5_mtmm}
\end{figure*}
\paragraph{From NAR to AR}
Diffusion-based dance generation models are typically non-autoregressive,  
making it difficult to generate long dance sequences efficiently and maintain temporal coherence.  
Extending to long-sequence or full-audio inference significantly increases computational cost  
and reduces output quality.
The Motion Temporal Mamba Module (MTMM) addresses this by incorporating autoregressive temporal context  
into the diffusion process.  
It leverages the Mamba state-space architecture to model long-range motion dependencies efficiently.  

As shown in Figure~\ref{fig:fig5_mtmm}, latent features from the previous segment (motion memory)  
are concatenated with the current segment's latents, followed by positional encoding.  
A bidirectional Mamba scan—consisting of forward and reversed sequence passes—processes this combined sequence.  
The output from both scans is merged, normalized, and split to yield the latent representation  
for the current motion segment, which is fed into the next generation step.  
This design enables smooth temporal transitions and coherent long-duration dance synthesis.

Given the latent motion sequence for the current segment  
$X_{n^\prime} \in \mathbb{R}^{T_{\mathrm{seq}} \times D_p}$  
(which is produced by the dual-layer Conformer integrating text and audio inside the previous denoising block)  
and the motion memory from the tail of the preceding segment  
$x_m \in \mathbb{R}^{T_m \times D_p}$  
(where $T_m$ denotes the number of tail frames),  
our autoregressive diffusion model aims to guide the generation of the next motion segment  
via temporal features from the previous segment.

Both $x_m$ and $X_{n^\prime}$ are processed in a front-end layer implemented as Conv1D-based MLPs  
to project features into a common dimension $C_m$.  
They are then concatenated along the temporal dimension,  
and positional encoding $PE \in \mathbb{R}^{(T_m+T_{\mathrm{seq}}) \times C_m}$  
is added to form the input $x_{\mathrm{pos}}$ for the bidirectional Mamba scan:
\begin{equation}
x_{\mathrm{pos}} = 
\mathrm{LN}\left(
    \mathrm{Concat}\left( f_\varepsilon(x_m),\, f_\theta(X_{n^\prime}),\, \mathrm{dim} = 1 \right)
    \oplus PE
\right)
\label{eq:mtmm_input}
\end{equation}

The MTMM performs a bidirectional scan:  
a forward SSM pass $\mathcal{M}_f(\cdot)$ on $x_{\mathrm{pos}}$ and  
a backward SSM pass $\mathcal{M}_b(\cdot)$ after sequence reversal.  
The backward output is then re-reversed and added element-wise to the forward result:
\begin{equation}
x_{\mathrm{mem}} =
\mathcal{M}_f(x_{\mathrm{pos}}) \oplus
\mathrm{Reverse}\left[ \mathcal{M}_b\left( \mathrm{Reverse}(x_{\mathrm{pos}}) \right) \right]
\label{eq:mtmm_scan}
\end{equation}
The combined features $x_{\mathrm{mem}}$ are normalized and split along the temporal axis  
into $x_h$ (current motion features) and the preceding segment’s portion.  
Residual connections and layer normalization are applied:
\begin{equation}
\_,\, x_h = \mathrm{Split}\left( \mathrm{LN}(x_{\mathrm{mem}}),\ [T_m, T_{\mathrm{seq}}],\, \mathrm{dim} = 1 \right)
\label{eq:mtmm_split}
\end{equation}
\begin{equation}
x_{\mathrm{res}^\prime} = \mathrm{LN}\left( X_{n^\prime} + \mathrm{LN}(x_h) \right)
\label{eq:mtmm_residual}
\end{equation}

Following the structure in Section~3.3,  
$x_{\mathrm{res}^\prime}$ is split into gated and filtered components $h_{g^\prime}$ and $h_{f^\prime}$,  
activated via $\tanh(\cdot)$ and $\sigma(\cdot)$,  
and fused via a linear projection $f_\delta(\cdot)$:  
\begin{equation}
h_{g^\prime}, h_{f^\prime} = \mathrm{Split}(x_{\mathrm{res}^\prime},\, 2,\, \mathrm{dim} = 2)
\label{eq:mtmm_gate_filter}
\end{equation}
\begin{equation}
X_{n-1}, h_{\mathrm{skip}} =
\mathrm{Split}\left(
    f_\delta\left( \tanh(h_{f^\prime}) \odot \sigma(h_{g^\prime}) \right)
\right)
\label{eq:mtmm_output}
\end{equation}
Here, $X_{n-1}$ is the processed latent motion for the next denoising step,  
and $h_{\mathrm{skip}}$ is passed through skip connections.

\begin{figure}[t]
    \centering
    \includegraphics[width=\linewidth]{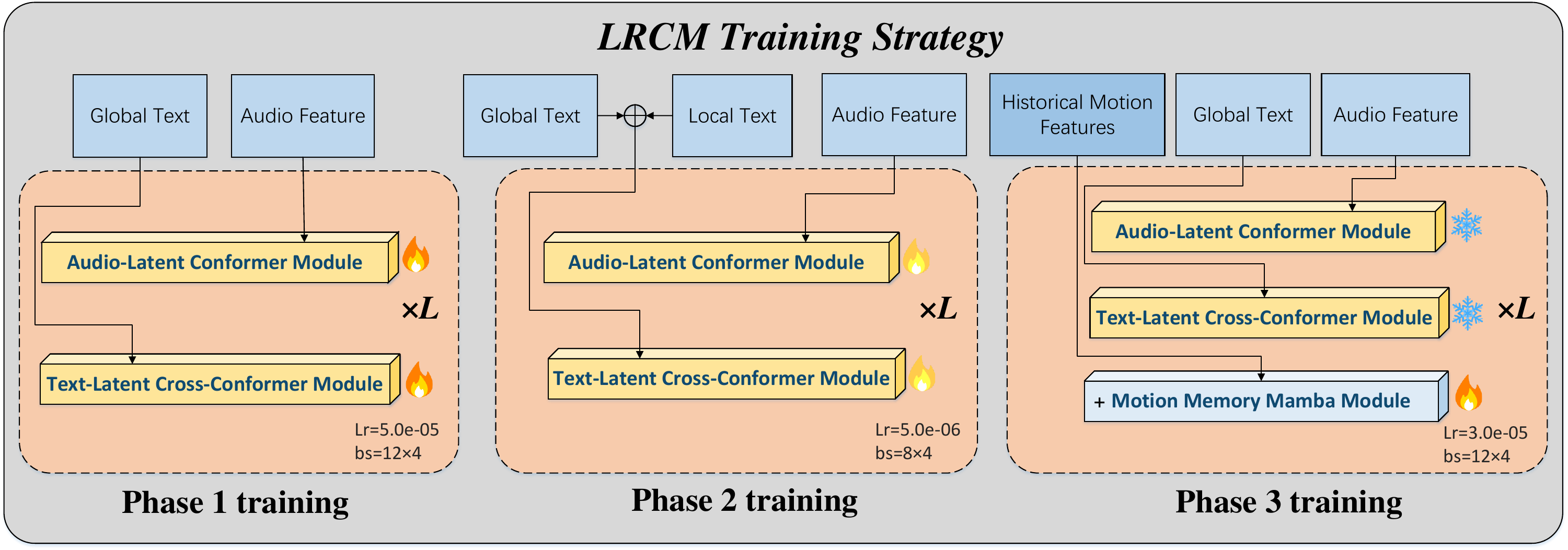}
    \vspace{-2.0em}
    \caption{Full LRCM model training strategy showing three-phase workflow and module freezing schedule.}
    \Description{Diagram illustrating Phase 1 (global text + audio), Phase 2 (add local text), and Phase 3 (enable MTMM with module freezing).}
    \label{fig:fig6_training_strategy}
\end{figure}

\paragraph{Training Strategy.}
For the multi-input autoregressive dance generation task,  
we adopt a three-phase training strategy (visualized in Figure~\ref{fig:fig6_training_strategy}):
\begin{itemize}
    \item \textbf{Phase 1}: Train the non-autoregressive backbone using only global text embeddings and audio embeddings.  
    A relatively high learning rate is used to strengthen global text guidance for overall motion generation.
    \item \textbf{Phase 2}: Fine-tune with both global and local text inputs alongside audio,  
    lowering the learning rate to improve alignment between fine-grained textual descriptions and motion details.
    \item \textbf{Phase 3}: Enable the Motion Temporal Mamba Module (MTMM) while freezing prior audio–text fusion layers,  
    focusing on temporal continuity and coherence in long-sequence dance synthesis.
\end{itemize}

\section{Experience}
\subsection{Experimental Setup}
\paragraph{Dataset.}
All experiments are conducted using the decoupled multimodal dance dataset  
constructed in Section~3.  
This dataset focuses on street dance styles and contains high-quality motion capture data,  
synchronized audio, and global/local textual annotations labeled by professional dancers.  
As summarized in Table~\ref{tab:dataset_stats},  
the extended Motorica Dance dataset exhibits a unique advantage in the richness of textual annotations,  
particularly in the number of local text segments.  
Detailed token statistics for each genre are provided in the Appendix~\ref{sec:dataset_detail}.  
For the scope of this work,  
we only perform decoupling and generalization on the Motorica Dance dataset.  
The dataset is split into dance clips,  
ensuring that no single clip appears in both the training and testing sets.  
This separation enables a reliable evaluation of the model’s generalization capability.

\begin{table}[ht]
\setlength{\abovecaptionskip}{3pt} % 标题和表格的距离
\setlength{\belowcaptionskip}{-2pt} % 标题和正文的距离(可选)
\caption{Statistics of the decoupled multimodal dance dataset.}
\label{tab:dataset_stats}
\centering
\begin{tabular}{lcccc}
\toprule
Genre       & Full Time & Global Tokens & Local Tokens & All Tokens \\
\midrule
Hiphop      & 84 mins   & 233           & 629          & 862 \\
Krumping    & 18 mins   & 62            & 437          & 499 \\
Popping     & 42 mins   & 124           & 528          & 652 \\
Locking     & 18 mins   & 40            & 154          & 194 \\
Jazz        & 52 mins   & 89            & 345          & 434 \\
Charleston  & 50 mins   & 134           & 465          & 599 \\
Tapping     & 11 mins   & 27            & 45           & 72 \\
\bottomrule
\end{tabular}
\end{table}

\paragraph{Evaluation Metrics.}
We assess generated dance quality using:  
\textbf{FID} ($\mathrm{FID}_k$: skeleton kinematics, $\mathrm{FID}_g$: geometric features),  
\textbf{Rhythm Alignment} (Beat Alignment Score, BAS),  
\textbf{Diversity} (mean pairwise Euclidean distance, $\mathrm{DIV}_k$: skeleton kinematics, $\mathrm{DIV}_g$: geometric features),  
and an improved \textbf{Freezing Analysis} with Adaptive PFF (PFF), Rhythmic Score (RS), and Length Regularity (LR).  
Formal definitions are given in the Appendix~\ref{sec:evaluation_metrics}.

\paragraph{Implementation Details.}
Our PyTorch model is trained on 4$\times$RTX 4090 (24\,GB),  
using the \texttt{diffusers} DDPM sampler (200 steps, 20 residual blocks).  
Phase~1: global text + audio, LR=$5.0{\times}10^{-5}$, bs=12, 100h.  
Phase~2: add local text, LR=$5.0{\times}10^{-6}$, bs=8, 5h.  
Phase~3: enable MTMM, freeze fusion layers, LR=$3.0{\times}10^{-5}$, bs=12, $\sim$20h.  
Optimizer: Adam, weight decay=$1.0{\times}10^{-4}$.  
0.05 probability Gaussian noise injection per input modality improves robustness.  
Final model size: $\sim$316M parameters.

\subsection{Quantitative Results}
\paragraph{Comparison Models.}
We compare our method against several mainstream architectures:  
VQ-VAE-based models (\textit{TM2D}~\cite{gong2023tm2d}), \textit{Dance-ba}~\cite{fan2025align}),  
diffusion-based models (\textit{GCDance}~\cite{liu2025gcdance}), \textit{Lodge}~\cite{li2024lodge}),  
and the baseline \textit{LDA}~\cite{alexanderson2023listen}.  
All baselines are trained using the proposed decoupled dance dataset format.  

The test set includes six full-length music tracks,  
covering all dance genres in Motorica Dance,  
with an average track length of approximately three minutes.  
Some baseline models only support genre-level input,  
which is encoded as a one-hot vector,  
and cannot process full textual descriptions.  
During evaluation, these models are trained and tested with genre one-hot input,  
while our model uses complete text prompts.  
Due to hardware constraints on RTX~4090 GPUs,  
most comparable models cannot support full-length inference without running out of memory.  
We therefore perform long-sequence evaluation by segmenting the tracks  
and stitching results.  
A summary of supported functionalities for each model is shown in Table~\ref{tab:model_capabilities}.

\begin{table}[ht]
\setlength{\abovecaptionskip}{3pt} % reduce caption-table gap
\caption{Model capability comparison.}
\label{tab:model_capabilities}
\centering
\resizebox{\columnwidth}{!}{
\begin{tabular}{lccccc}
\toprule
Method   & Venue & Audio & Text & Long Audio & Memory Autoreg. \\
\midrule
GCDance  & Arxiv25 & \checkmark & --          & --          & --          \\
Danceba  & ICCV25 & \checkmark & --          & --          & --          \\
Lodge    & CVPR24 & \checkmark & --          & \checkmark  & --          \\
TM2D     & ICCV23 & \checkmark & \checkmark  & --          & --          \\
LDA      & SIG23 & \checkmark & --          & --          & --          \\
Ours     & - & \checkmark & \checkmark  & \checkmark  & \checkmark  \\
\bottomrule
\end{tabular}
}
\end{table}
\paragraph{Objective Metric Comparison.}
We compare our method against state-of-the-art dance motion generation models  
on the decoupled dataset using standard quantitative metrics  
(Table~\ref{tab:metrics_comparison}).  
Our method achieves \textbf{lowest} FID scores for both kinematic ($\mathrm{FID}_k$) and geometric ($\mathrm{FID}_g$) motion quality,  
outperforming all prior works by over 35\% relative improvement.  
Compared to the baseline LDA, our method yields $7.5\%$ and $10.8\%$ improvements  
on $\mathrm{FID}_k$ and $\mathrm{FID}_g$ respectively.  

In motion diversity, our approach records the highest scores in both kinematic and geometric diversity,  
exceeding all competitors by more than 42\%.  
Unlike category label encoding, full textual descriptions do not impose overly strict constraints on the diversity of generated dances,  
allowing the model to produce a broader range of motion styles while maintaining high quality.  
This flexibility yields a slight advantage over the ground truth (GT) data in diversity metrics.  
For Beat Alignment Score (BAS), Lodge attains the top result,  
surpassing ours by a small margin ($+2.2\%$),  
while our method ranks second ahead of the baseline.

Overall, the results indicate that our framework delivers superior motion quality and diversity,  
while maintaining competitive rhythm alignment performance in long-sequence synthesis.

\begin{table}[ht]
\setlength{\abovecaptionskip}{3pt}
\caption{Comparison with state-of-the-art methods on the decoupled dataset.  
GT denotes ground truth. Bold indicates best, underlined second-best.  
$\downarrow$: lower is better, $\uparrow$: higher is better.}
\label{tab:metrics_comparison}
\centering
\begin{tabular}{lccccc}
\toprule
Method & $\mathrm{FID}_k\downarrow$ & $\mathrm{FID}_g\downarrow$ & DIV$_k\uparrow$ & DIV$_g\uparrow$ & BAS$\uparrow$ \\
\midrule
GT       & --       & --       & 62.47 & 85.70 & 0.2941 \\
TM2D     & 1445.59  & 8708.70  & 36.97 & 15.62 & 0.1041 \\
Lodge    & 346.31   & 9169.34  & 45.36 & 23.26 & \textbf{0.2898} \\
Danceba  & 2339.61  & 9789.57  & 2.11  & 0.97  & 0.2574 \\
GCDance  & 906.04   & 9949.49  & 30.52 & 14.29 & 0.2187 \\
LDA      & \underline{275.57} & \underline{3556.10} & \underline{61.63} & \underline{70.65} & 0.2710 \\
\midrule
Ours     & \textbf{256.42}    & \textbf{3208.32}    & \textbf{87.83}    & \textbf{86.38}    & \underline{0.2835} \\
\bottomrule
\end{tabular}
\end{table}

\paragraph{Freezing and User Study Results.}
To evaluate whether generated motions exhibit unintended freezing,  
we use a motion freezing rate metric.  
Results are shown in Table~\ref{tab:freezing_results}.  

From a numerical perspective, the ground truth is not necessarily the optimal score for freezing proportion;  
we evaluate each algorithm by its proximity to GT values.  
Our framework surpasses the baseline LDA and all other methods in \textbf{Adaptive PFF},  
achieving the closest score to GT.  
For \textbf{Rhythmic Score} (RS), our method ties with LDA for first place,  
and ranks second for \textbf{Length Regularity} (LR).  
Overall, our generated motions maintain freezing statistics closest to the GT distribution.

Appropriate freezing at certain beats is desirable in dance generation.  
Although some algorithms have lower numerical freezing rates,  
they may still exhibit freezing subjectively;  
this will be further addressed in Section~4.3 with qualitative frame-by-frame analysis.

\begin{table}[ht]
\setlength{\abovecaptionskip}{3pt}
\caption{Freezing rate and user study results.  
Bold: best, Underlined: second-best.  
Values indicate deviation from GT ($+/-$ higher/lower than GT).}
\label{tab:freezing_results}
\centering
\begin{tabular}{lccc}
\toprule
Method   & PFF ($\Delta$) & RS ($\Delta$)  & LR ($\Delta$) \\
\midrule
GT       & 0.0748         & 0.9875         & 0.0768 \\
TM2D     & +0.0596        & -0.0390        & -0.0259 \\
Lodge    & -0.0335        & \underline{+0.0018} & +0.2654 \\
Danceba  & -0.0706        & +0.0039        & +0.7181 \\
GCDance  & -0.0534        & -0.0041        & +0.4105 \\
LDA      & \underline{+0.0248} & \textbf{+0.0005} & \textbf{-0.0135} \\
\midrule
Ours     & \textbf{+0.0143}    & \textbf{-0.0005} & \underline{+0.0244} \\
\bottomrule
\end{tabular}
\end{table}

\subsection{Qualitative Results}

Traditional qualitative evaluation of dance generation examines motion smoothness, music alignment, and style consistency.  
For our proposed architecture, we additionally assess the subjective influence of textual descriptions on the generated motion  
and evaluate clip-to-clip transitions under autoregressive inference for long audio sequences.

\begin{figure}[t]
    \centering
    \includegraphics[width=\linewidth]{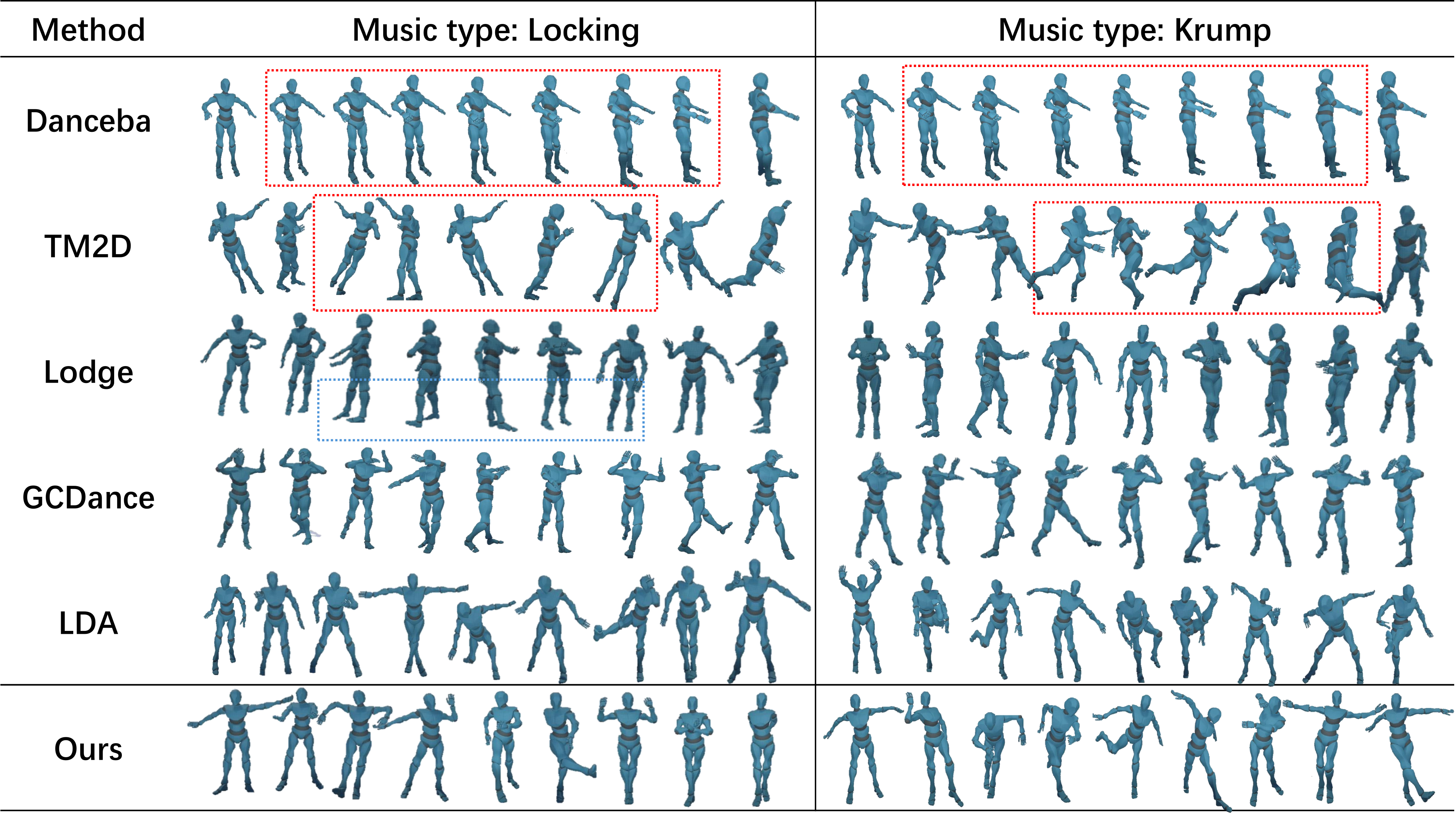}
    \vspace{-2.0em}
    \caption{Qualitative comparison of generated results across methods.  
    Red boxes: abnormal overall motion; blue boxes: abnormal local motion.}
    \label{fig:qual_results1}
    \Description{Comparison of generated dance frames from different models, with two sample music tracks.}
\end{figure}

Figure~\ref{fig:qual_results1} compares generation results from various SOTA methods using two sample music tracks.  
For encoder–decoder architectures with non-diffusion such as VAE and VQ-VAE,  
which perform well in video-based 3D reconstruction datasets,  
training and generalization to traditional motion-capture data remains challenging.  
Models like Danceba tend to collapse to a \textit{mean pose} in generation,  
visually appearing frozen; TM2D yields mean-pose states with rotational artifacts;  
GCdance produces extremely uniform motion patterns.  
Diffusion-based approaches such as Lodge and LDA avoid full freezing on our dataset,  
but still exhibit varying degrees of motion blur, semantic–style inconsistency,  
and fragmented autoregressive transitions.  
In contrast, our model demonstrates high motion quality in long-sequence autoregressive inference  
when conditioned on both audio and text.

\begin{figure}[t]
    \centering
    \includegraphics[width=\linewidth]{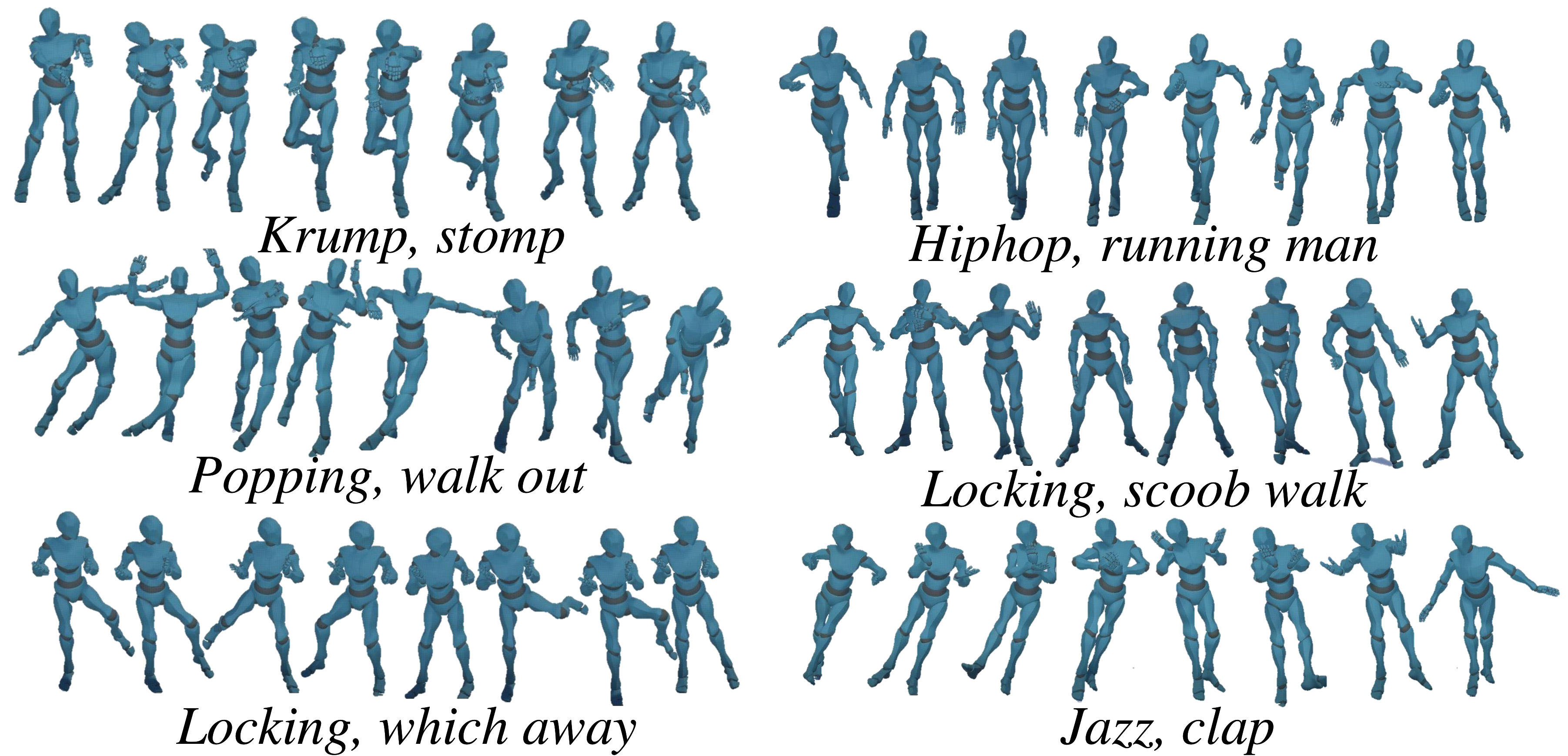}
    \vspace{-2.0em}
    \caption{Phase~2 results after fine-tuning with local text.  
    Generated motions from diverse textual prompts.}
    \label{fig:qual_results2}
    \Description{Generated dance frames from various text prompts, demonstrating alignment with described actions.}
\end{figure}

We further examine our model’s text–audio guided generation by providing diverse textual instructions.  
Figure~\ref{fig:qual_results2} shows full-audio inference results for several motions:  
(a) \textit{Krump, stomp} — clear leg lift and step;  
(b) \textit{Hiphop, running man} — running motion and arm swings;  
(c) \textit{Popping, walk out} — stepping with clean upper body posture;  
(d) \textit{Locking, scoob walk} — leg kick followed by knee joint locking/unlocking;  
(e) \textit{Locking, which away} and \textit{Jazz, clap} — high completion score with distinct hand/leg movements.  

While some motions lack fine-detail reproduction,  
the core action aligns well with the textual and musical cues for common dance styles.

\paragraph{Effect of MTMM on Autoregressive Transitions.}
We study the impact of the MTMM module (Sec.~3.4) on clip-to-clip autoregressive generation.  
Each clip consists of 900 frames, and we focus on transitions at 30-second boundaries.

\begin{figure}[t]
    \centering
    \includegraphics[width=\linewidth]{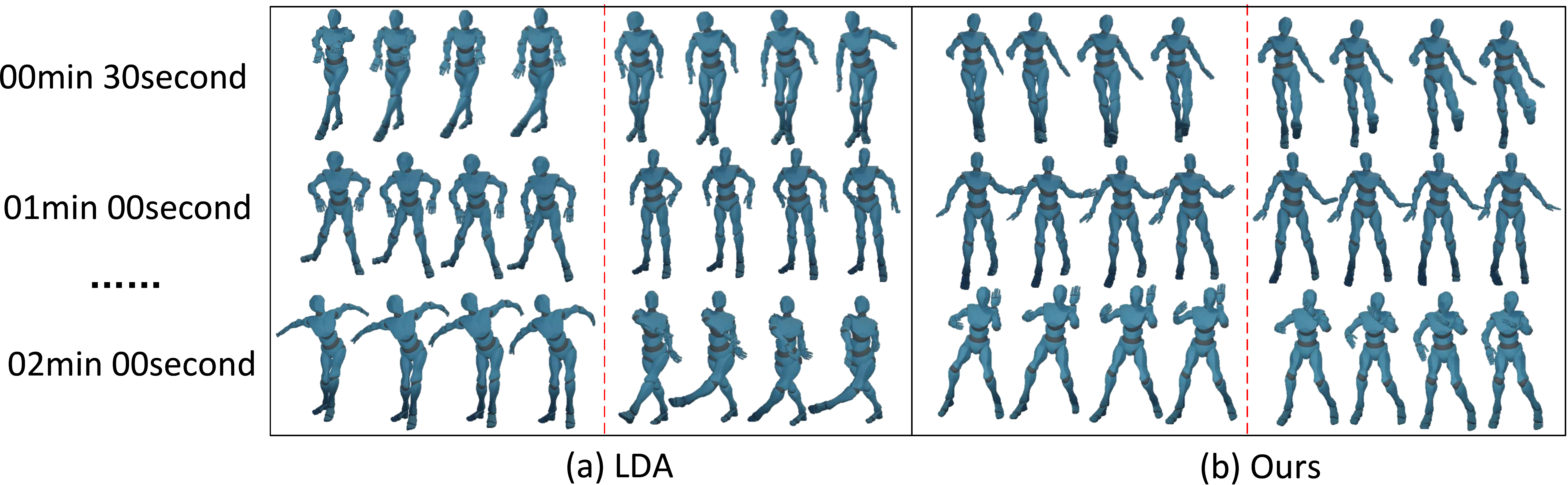}
    \vspace{-2.0em}
    \caption{MTMM transition comparison in Phase~3 training.  
    (a) Non-autoregressive LDA; (b) Ours with MTMM.  
    Red dashed lines indicate the clip boundary.}
    \label{fig:mtmm_transition}
    \Description{Comparison of dance frames around clip transitions, showing smoother transitions with MTMM.}
\end{figure}

Figure~\ref{fig:mtmm_transition} compares frame sequences before and after transitions for the non-autoregressive LDA  
and our MTMM-enhanced model.  
The MTMM enables smooth temporal blending between clips,  
producing bolder motions with greater amplitude and richer joint rotations  
within and across clips—demonstrating its benefit for both intra-clip motion quality and inter-clip continuity.

\subsection{Ablation Studies}

We conduct ablation experiments to evaluate the contribution of key components in our framework—text conditioning, the MTMM, and local text guidance.  
Table~\ref{tab:ablation_results} summarizes the results.

The full LRCM model consistently outperforms all ablated variants across all metrics.  
Removing textual prompts (\texttt{w/o text/"..."}) or the MTMM leads to significant drops in motion quality, diversity, and BAS.  
Notably, \texttt{w/o local\_text} exhibits performance closer to the ground truth (GT) in certain geometric attributes,  
suggesting that local text mainly enhances fine-grained semantic control rather than coarse geometric fidelity.  
In terms of semantic control, coarse-grained global text can preserve geometric properties similar to the GT,  
whereas local text provides finer control over motion semantics, though in GT-alignment metrics it may perform slightly worse than using only global text.

\begin{table}[ht]
\setlength{\abovecaptionskip}{3pt}
\caption{Ablation results. Bold: best performance, Underlined: second-best. $\downarrow$: lower is better, $\uparrow$: higher is better.}
\label{tab:ablation_results}
\centering
\begin{tabular}{lccccc}
\toprule
Method & FID$_k\downarrow$ & FID$_g\downarrow$ & DIV$_k\uparrow$ & DIV$_g\uparrow$ & BAS$\uparrow$ \\
\midrule
w/o text/"..." & \underline{294.10} & 3376.91         & 68.66                   & 59.96                   & 0.2739 \\
w/o MTMM       & 392.58             & 3935.48         & 41.89                   & 48.88                   & 0.2068 \\
w/o local\_text & 448.19             & \textbf{2496.13} & \underline{83.14}       & \textbf{87.67}          & \textbf{0.2981} \\
Full LRCM      & \textbf{256.42}    & \underline{3208.32} & \textbf{87.83}        & \underline{86.38}       & \underline{0.2835} \\
\bottomrule
\end{tabular}
\end{table}

\paragraph{Noise Scheduler Analysis.}
We also investigate the effect of the noise scheduler configuration.  
Using the baseline LDA with DDPM parameters $\beta \in [0.01, 0.7]$ and a linear schedule with 150 diffusion steps,  
we observe that larger $\beta$ values prematurely push the model into high-noise regimes (around step~50),  
where little useful denoising information can be learned, leading to wasted computation.

\begin{figure}[t]
    \centering
    \includegraphics[width=\linewidth]{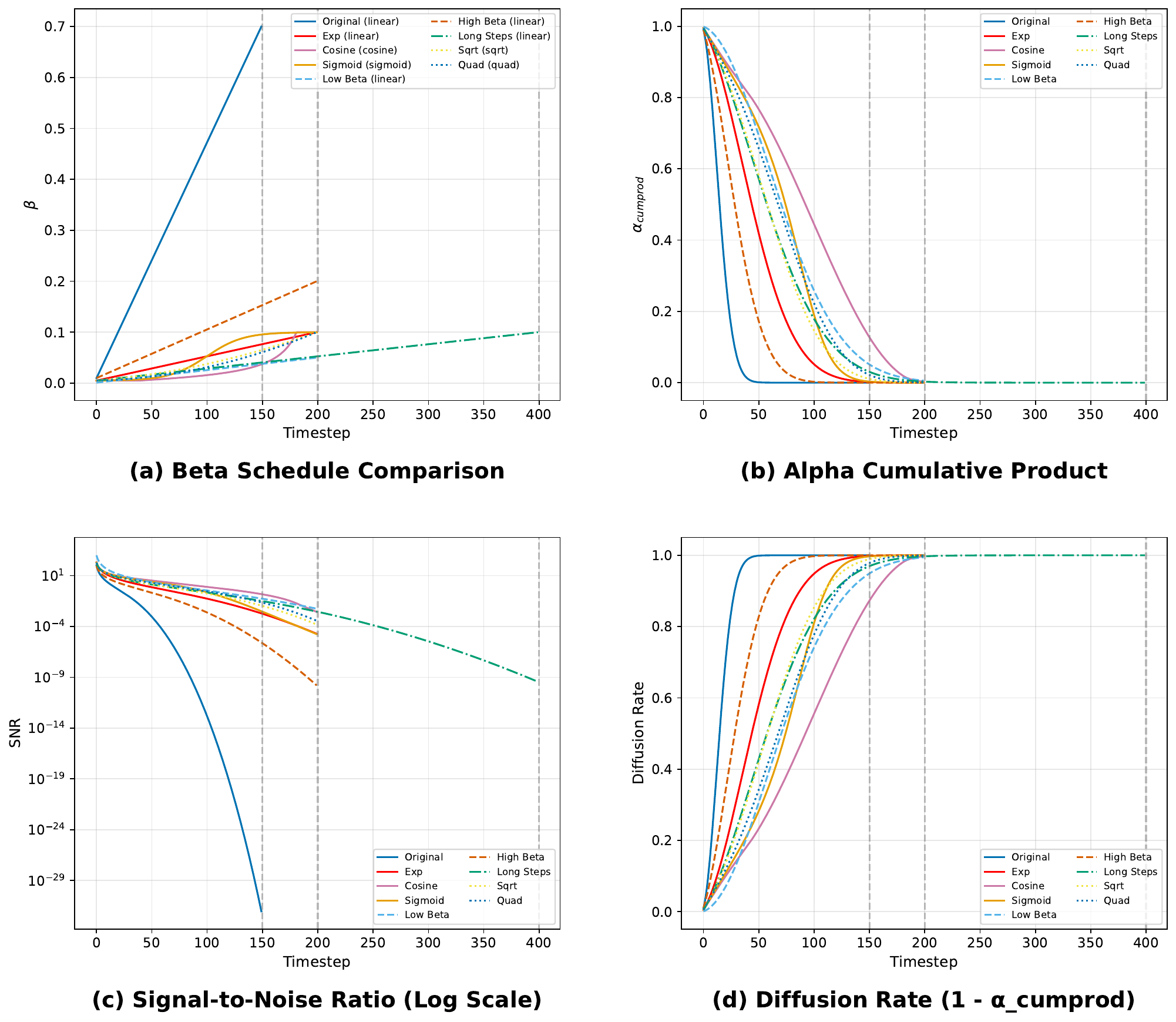}
    \caption{Noise scheduler $\beta$ and $\alpha$ curves.  
    Blue: baseline LDA configuration; Red: experimental setting used in this work;  
    other curves: alternative reference.}
    \Description{Graph comparing noise scheduler curves for different $\beta$ and $\alpha$ configurations.  
    Blue denotes the baseline LDA schedule, red indicates the experimental configuration used in our study,  
    and additional curves correspond to alternative scheduling strategies.}
    \label{fig:noise_scheduler}
\end{figure}

In our configuration, we set $\beta \in [0.005, 0.1]$ with $T=200$ diffusion steps,  
delaying the transition to high-noise steps.  
This adjustment reduces motion jitter in inference and improves utilization of high-noise training phases.
Figure~\ref{fig:noise_scheduler} plots the $\beta$ and $\alpha$ curves for the original LDA setting (blue)  
and our modified configuration; additional curves are provided for reference.

\section{Conclusion and Future Work}

In this paper, we presented a novel framework for multimodal dance generation.  
Recognizing the importance of semantic textual information in dance studies,  
we proposed a semantic-decoupling paradigm based on the Motorica Dance dataset.  
The dataset was disentangled into motion, audio, and global/local textual modalities  
across seven dance genres, enabling fine-grained pairing of global and temporal segment descriptions.

We designed an audio–text multimodal diffusion model,  
leveraging Conformer and Cross-Conformer architectures for effective 1D sequence modeling  
in a non-autoregressive setting.  
For long-duration autoregressive generation,  
we introduced the Motion Temporal Mamba Module (MTMM) based on the Mamba state-space architecture,  
bridging non-autoregressive and autoregressive paradigms.  

Our complete LRCM framework was developed as an exploration of text-guided and long-sequence dance generation,  
and demonstrated strong performance across multiple objective metrics and qualitative evaluations.  
The experiments extend the study of semantic-decoupling schemes, multimodal conditioning,  
and autoregressive temporal modeling, providing evidence of their potential benefits  
for complex and sustained motion synthesis tasks.

\paragraph{Future Work.}
The proposed dataset decoupling approach can be further refined and expanded  
to include richer annotations and additional modalities.  
Our current exploration of text input in an autoregressive framework  
represents an initial step;  
future work will focus on improving fine-detail synthesis,  
enhancing long-sequence generation stability,  
and extending the architecture for broader genre coverage and cross-domain motion tasks
such as inpainting and editing.

\bibliographystyle{ACM-Reference-Format}
\bibliography{cited_references}

\appendix

\section{Large Language Model Fine-tuning Details}
\label{sec:llm_finetune}

To enable professional-level textual guidance for dance motion generation,  
we fine-tuned a large language model (LLM) to serve as a domain-adapted text preprocessor.  
The model was trained to generate structured and semantically rich dance descriptions---including \textit{global style tokens} and \textit{local movement tokens}---from natural language prompts.  
In our work, we use the \texttt{GLM-4-FLASH} model~\cite{glm2024chatglm} as the base LLM,  
chosen for its lightweight architecture and high efficiency within the overall framework.

\subsection{Fine-tuning Corpus}
The fine-tuning dataset was manually constructed and stored in a JSONL conversational format,  
where each entry contained a multi-turn dialogue between:
\begin{itemize}
    \item \textbf{System}: Instructing the model to behave as a \textit{professional dance artist} capable of choreographing dances based on user prompts.
    \item \textbf{User}: Providing informal or varied natural language prompts,  
          often mixed with dance move names or style descriptions (e.g., \texttt{running man, step torch, heel toe}).
    \item \textbf{Assistant}: Responding with a two-part structured annotation:
    \begin{enumerate}
        \item \textit{Dance style characteristics} — a high-level global style descriptor (e.g., genre, expressiveness, moving range, key stylistic elements).
        \item \textit{Dance movement sequence} — a temporally segmented list of local action tokens with start/end timestamps.
    \end{enumerate}
\end{itemize}

\subsection{Annotation Format}
Each assistant response is formatted as:
\begin{verbatim}
Dance style characteristics: <global tokens>
Dance movement sequence:
- HH:MM:SS.sss - HH:MM:SS.sss: <local tokens>
...
\end{verbatim}
where:
\begin{itemize}
    \item \textbf{Global tokens} capture coarse-grained style semantics (e.g., \texttt{popping, extrovert, medium moving range, isolation}).
    \item \textbf{Local tokens} specify temporally localized actions (e.g., \texttt{walk out, foot movement}, \texttt{head isolation}, \texttt{spin}).
\end{itemize}

\subsection{Purpose and Integration}
The fine-tuned LLM is integrated into the LRCM pipeline as a \textbf{textual semantic expansion module} before the CLIP-based text encoder.  
It normalizes unstructured, noisy, or incomplete user prompts into consistent global/local token pairs,  
improving the precision of multimodal conditioning for the diffusion model.  
This step was crucial to handling non-expert or freestyle text inputs while retaining professional choreography quality.

\section{Dataset Detail}
\label{sec:dataset_detail}
\begin{figure}[t]
    \centering
    \includegraphics[width=\linewidth]{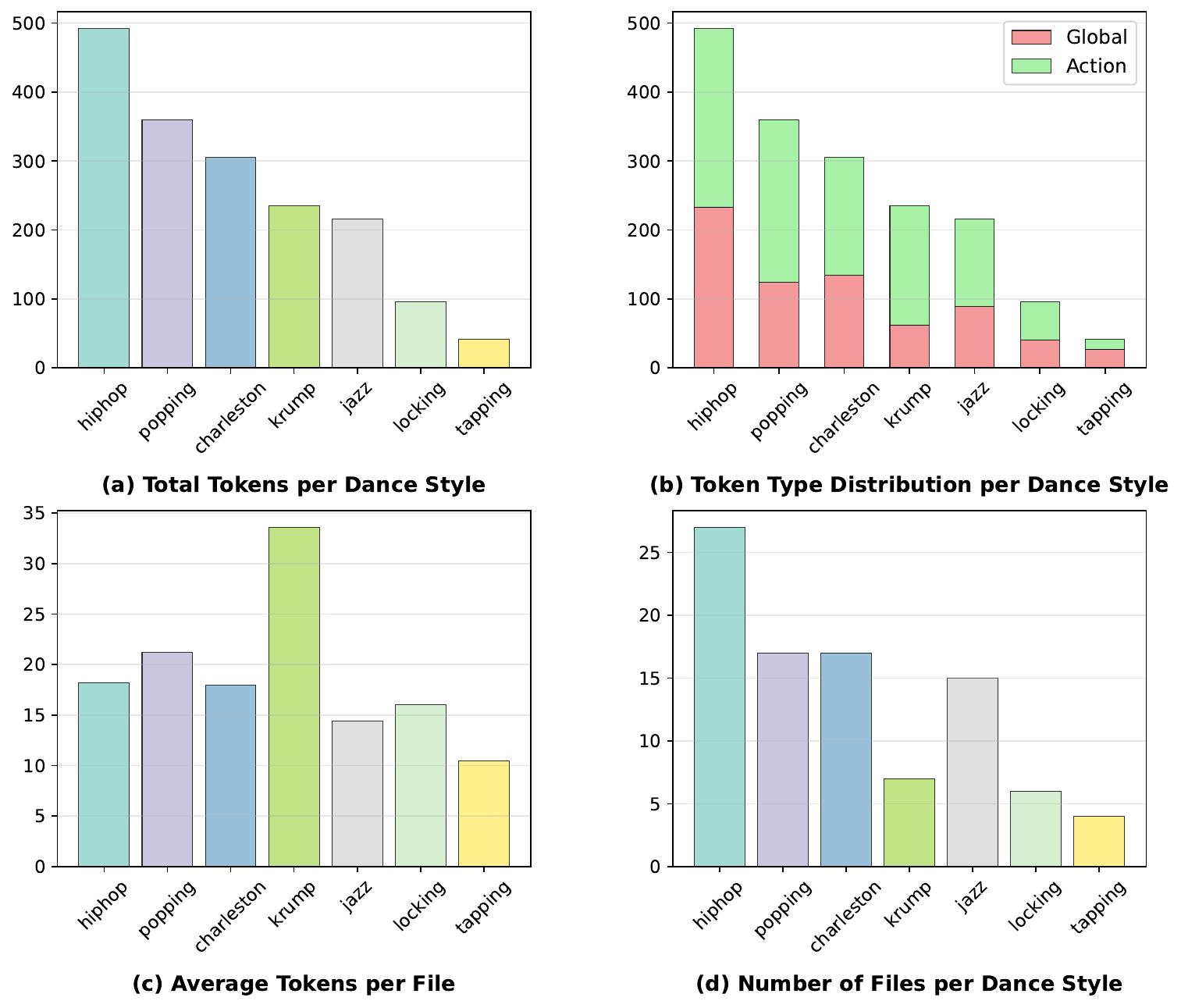}
    \caption{Token statistics for all dance styles in the Motorica Dance dataset.}
    \Description{
        (a) Total number of tokens per dance style;  
        (b) Global vs. local token counts;  
        (c) Average text descriptions per clip by style;  
        (d) Number of clips per style.
    }
    \label{fig:dataset_tokens_stats}
\end{figure}

\begin{figure}[t]
    \centering
    \includegraphics[width=\linewidth]{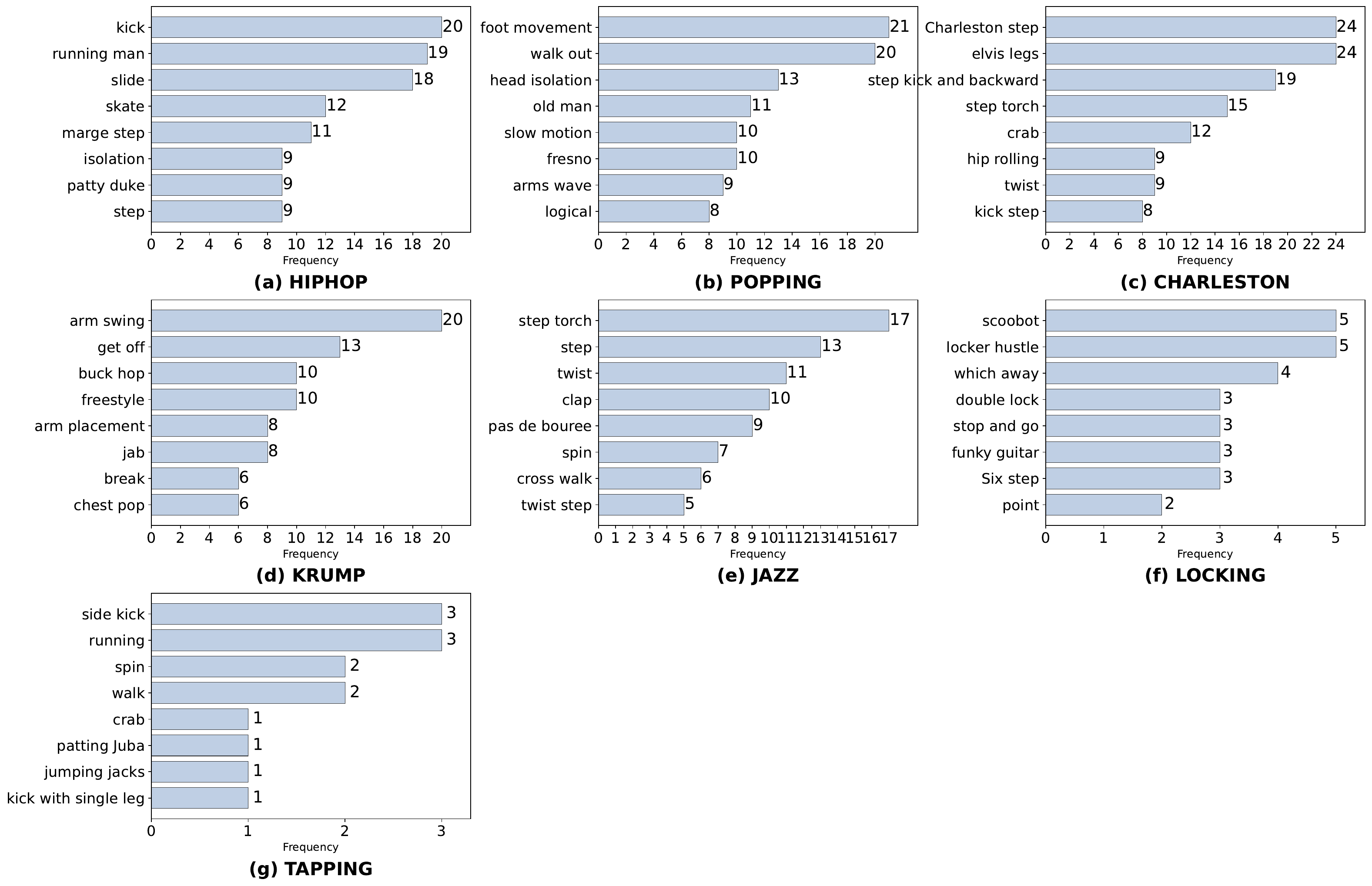}
    \caption{Top eight most frequent semantic tokens for each dance style.}
    \Description{Bar charts showing the eight most common tokens for each dance style.}
    \label{fig:style_common_tokens}
\end{figure}

\begin{figure}[t]
    \centering
    \includegraphics[width=\linewidth]{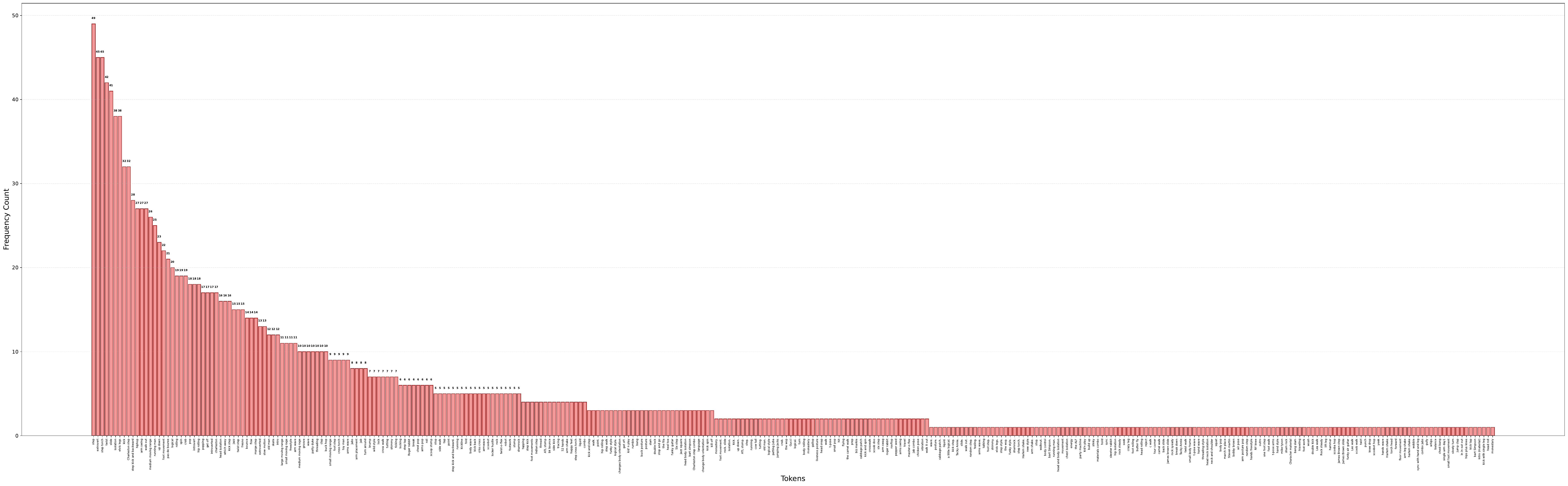}
    \caption{Summary of all dance action tokens across the dataset.}
    \Description{Bar chart summarizing the frequency of all dance action tokens in the dataset.}
    \label{fig:all_tokens_summary}
\end{figure}

In this work, we conduct professional motion–semantic decoupling  
for each of the seven dance styles in the Motorica Dance dataset.  

Figure~\ref{fig:dataset_tokens_stats}(a) shows the total number of tokens per style.  
Figures~\ref{fig:dataset_tokens_stats}(b)--(d) illustrate global vs. local token counts,  
average text descriptions per dance clip by style, and the number of clips per style, respectively.  

The variation across styles reflects the nature of the dance types.  
Figure~\ref{fig:style_common_tokens} lists the most frequent token entries for each style, 
while Figure~\ref{fig:all_tokens_summary} provides the global summary of all dance action tokens, 
revealing style-specific semantics:

\begin{itemize}
    \item \textbf{Hiphop}: Dominated by groove-related terms such as \texttt{up down} and \texttt{groove},  
    repeated throughout the style. Signature actions like \texttt{running man}, \texttt{step}, and footwork entries are also prevalent.
    \item \textbf{Krump}: Prominent hand actions (\texttt{arm swing}, \texttt{jab}) and strong footwork such as \texttt{stomp},  
    often spanning the entire dance sequence.
    \item \textbf{Popping}: The term \texttt{pop} appears in every global token set;  
    fundamental steps like \texttt{walk out} and \texttt{toy man} are linked to \texttt{foot movement}.  
    Logical-style elements like \texttt{tutting} and \texttt{wave} are frequent in the global descriptions.
    \item \textbf{Locking}: Dominated by point-oriented actions like \texttt{point lock}, generally short in duration.
    \item \textbf{Jazz}, \textbf{Charleston}, \textbf{Tapping}: Belonging to the jazz family,  
    featuring twist movements (\texttt{twist}, \texttt{torch}) and various step patterns,  
    distinct from street dance styles.
\end{itemize}

\section{Evaluation Metrics}
\label{sec:evaluation_metrics}
To comprehensively evaluate dance generation performance,  
we adopt several quantitative metrics covering motion quality, rhythm synchronization, diversity, and freezing phenomena.  
Formal definitions are provided below.

\subsection{Motion Quality – Fréchet Inception Distance (FID)}
FID measures the distributional difference between generated and real motions in the embedded feature space,  
assuming both follow Gaussian distributions:
\begin{equation}
\mathrm{FID} = \| \mu_r - \mu_g \|_2^2 + \mathrm{Tr} \left( \Sigma_r + \Sigma_g - 2 (\Sigma_r \Sigma_g)^{1/2} \right)
\end{equation}
where $\mu_r, \Sigma_r$ are the mean and covariance of real motion features,  
and $\mu_g, \Sigma_g$ are from generated motions.  
$\mathrm{FID}_k$ is computed from kinematic features (velocity, acceleration) derived from joint positions,  
and $\mathrm{FID}_g$ from geometric features (mean, standard deviation) of joint rotations.

\subsection{Beat Alignment Score (BAS)}
BAS evaluates synchronization between generated motion beats and musical beats,  
combining \textit{beat coverage} and \textit{beat alignment}:
\begin{equation}
\mathrm{BAS} =
\frac{1}{N_{\mathrm{beats}}}
\sum_{b \in \mathrm{MusicBeats}}
\delta_b \cdot \exp\left(-\frac{ \min_{m \in \mathrm{MotionBeats}}(m-b)^2 }{2\sigma^2} \right)
\end{equation}
where $N_{\mathrm{beats}}$ is the total number of music beats, $b$ is a beat index,  
$m$ is a generated motion beat frame, and $\delta_b=1$ if $b$ is covered by a motion keyframe, otherwise $0$.  
$\sigma$ denotes temporal tolerance (set to 9 frames in implementation).
\subsection{Diversity Score (DIV)}
Motion diversity is measured as the average Euclidean distance between pairs of generated motion feature vectors:
\begin{equation}
\mathrm{DIV} =
\frac{1}{K} \sum_{i=1}^K \| \mathbf{x}_{p_i} - \mathbf{x}_{q_i} \|_2
\end{equation}
where $K$ is the number of randomly sampled motion pairs,  
and $\mathbf{x}_{p_i}, \mathbf{x}_{q_i}$ are their corresponding feature vectors.

\subsection{Improved Freezing Metrics}
These metrics assess the occurrence and regularity of freezing in generated motion.
\paragraph{Adaptive Freezing Proportion (Adaptive PFF)}
\begin{equation}
\mathrm{Adaptive\;PFF} =
\frac{F_{\mathrm{adaptive}}}{T}
\end{equation}
where $F_{\mathrm{adaptive}}$ is the number of frames whose velocity falls below $\theta_v$  
for a duration in $[t_{\min}, t_{\max}]$,  
and $\theta_v=\mathrm{Percentile}_p(v_t)$ is the $p$th percentile threshold of the velocity distribution ($p=25\%$ by default).  
$T$ is the total number of motion frames.
\paragraph{Rhythmic Score (RS)}
\begin{equation}
\mathrm{RS} =
\frac{1}{N_b} \sum_{b=1}^{N_b}
\exp\left(-\frac{\min_{f \in F} | f - b |^2}{2\sigma^2} \right)
\end{equation}
where $F$ is the set of freeze points,  
$b$ is an expected beat location, and $\sigma$ denotes beat tolerance (default 30 frames).
\paragraph{Length Regularity (LR)}
\begin{equation}
\mathrm{LR} = \frac{1}{1 + \mathrm{Std}(L)}
\end{equation}
where $L$ is the set of durations of freezing segments, and $\mathrm{Std}(\cdot)$ is the standard deviation.

\end{document}